\documentclass{pasj00}
\draft

\newcommand{\ergs}{{\rm ergs}\ {\rm s}^{-1}}
\newcommand{\ergcms}{{\rm ergs}\ {\rm cm}^{-2}\ {\rm s}^{-1}}

\begin{document}
\SetRunningHead{Dogiel et al.}{Thermal and non-thermal hard X-ray
emission from the Galactic center}
\Received{2000/12/31}
\Accepted{2001/01/01}
\title{
 Origin of Thermal and Non-Thermal Hard X-ray Emission
from the Galactic Center}

\author{Vladimir \textsc{Dogiel}$^{1,2}$, Dmitrii {\sc Chernyshov}$^{2,3}$,
Takayuki {\sc Yuasa}$^4$, Dmitrii {\sc
Prokhorov}$^{3,5}$,  Kwong-Sang \textsc{Cheng}$^6$,   Aya {\sc
Bamba}$^{1}$, Hajime {\sc Inoue}$^{1}$, Chung-Ming
 {\sc Ko}$^7$, Motohide {\sc Kokubun}$^{1}$, Yoshitomo  {\sc
Maeda}$^{1}$, Kazuhisa {\sc Mitsuda}$^{1}$, Kazuhiro {\sc
Nakazawa}$^4$, and Noriko Y. {\sc Yamasaki}$^1$}
 \affil{$^1$Institute of Space and Astronautical
Science, 3-1-1, Yoshinodai, Sagamihara, Kanagawa, 229-8510, Japan}
  \affil{$^2$P.N.Lebedev Institute, Leninskii
pr, 53, 119991 Moscow, Russia, dogiel@lpi.ru}
\affil{$^3$Moscow Institute of Physics and Technology, Institutskii lane, 141700 
Moscow Region, Dolgoprudnii, Russia}
\affil{$^4$Department of
Physics, School of Science, The University of Tokyo, 7-3-1 Hongo,
Bunkyo-ku, Tokyo 113-0033}
 \affil {$^5$Institut d'Astrophysique de
Paris, CNRS, UMR 7095, Universit\'{e} Pierre et Marie Curie, 98bis
Bd Arago, F-75014 Paris, France} \affil{$^6$Department of Physics,
University of Hong Kong, Pokfulam Road, Hong Kong, China}
\affil{$^7$Institute of Astronomy, National Central University,
Jhongli 320, Taiwan}

%


\maketitle

\begin{abstract}
We analyse new results of {\it Chandra} and Suzaku which
found a flux of hard X-ray emission from the compact region around
Sgr A$^\ast$ (r$\sim 100$ pc). We suppose that this emission is
generated by accretion processes onto the central supermassive
blackhole when an unbounded part of captured stars obtains an
additional momentum. As a result a flux of subrelativistic protons
is generated near the Galactic center which  heats the background
plasma up to temperatures about $6-10$~keV and produces by inverse
bremsstrahlung a  flux of non-thermal X-ray emission in the energy
range above 10~keV.
\end{abstract}

\section{Introduction}

The X-ray emission of the order of $10^{38}$ erg s$^{-1}$ from the
Galactic ridge (GR) in the energy range above 1 keV was detected
almost forty years ago \citep{bleach} but its origin remains
unknown till now. Two possibilities are debated either this
emission is really diffuse \citep{ebi1, ebi2} or it is due to an
accumulative effect of faint X-ray sources \citep{rev1}. In the
case of diffuse origin serious energetic problems arise if this
emission is produced by non-thermal bremsstrahlung of high energy
electrons, since the required power of  sources of charged
particles in the disk should exceed $10^{42}$ erg s$^{-1}$, i.e.
higher than the power of supernovae  in the Galaxy
\citep{ski1,val,dog1}. However, this energy problem can be solved
if the particles are in-situ accelerated from background plasma
\citep{dog2,dog3}.

An alternative explanation of emission from GR was developed by
\citet{rev1} who presented recently essential arguments in favor
of the idea that the Galactic ridge emission was due to cumulative
emission of faint discrete X-ray sources. Though this
interpretation  is not completely proved at present, it appears to
be plausible. \citet{rev1} showed that the $3-20$~keV map of GRXE as
well as the 6.7 keV iron line  distribution in the disk closely
follow
 the near-infrared
(3.5 $\mu$) brightness distribution which traces the galactic
stellar mass distribution. This proportionality is the same in the
disk and in the bulge. From recent Suzaku observations it
was concluded  that the soft X-ray disk emission in the range 0.4
to 1 keV from the latitudes $<\timeform{2D}$ originated also from  faint
dM stars \citep{masui}.

Very decisive analysis about the origin of the Galactic ridge
emission was provided recently by \citet{rev09}. From the Chandra
data  they showed that most ($\sim 88$\%) of the ridge emission is
clearly explained by dim and numerous point sources.  Therefore,
at least in the ridge emission, accreting white dwarfs and active
coronal binaries are considered to be main emitters.

 The situation in the Galactic center (hereafter
GC) may be quite different. The GC region has been observed by
X-ray experiments flown for almost 30 years (see \cite{watson,
kawai}). Latter {\it Ginga} made a remarkable measurements of the
spectra in this region \citep{koya89,yama90}. Though as in the
case of GRXE, observations show there a significant contribution
of discrete sources with luminosity $L_{2-10~\rm{keV}}>10^{31}$
erg s$^{-1}$ which contribute from 20\% to 40\% of the total flux
\citep{muno,rev2}, the origin of the rest of the flux is still
unknown and  essential distinctions were found between flux
characteristics
 from  GRXE and GC. First of all, the
 GC emission is seen as a completely separated
 spherical region around Sgr-A$^\ast$ whose radius is about $100 -
 200$~pc \citep{muno,koya2}. The plasma temperature there is higher than
in other part of the Galactic disk. Secondly, the ratios of 6.9 to
6.7 keV lines (which
 traces the plasma temperature) and 6.4 to 6.7 keV lines are higher in
GC
 than in GR, while in GR this ratio is almost constant along the
  plane \citep{yama}.  Thirdly,  unlike the above-mentioned correlation of \citet{rev1} the X-ray source
 distribution derived from {\it Chandra} deep exposure of the $
<0.3^\circ$-radius central region does not show  any correlation
with the distribution of
 6.7 keV line  \citep{koya2}.  Therefore, they  concluded that the integrated
flux of point sources contributed a rather small fraction of the
total flux of GC X-rays and the major of emission from there is
diffuse.

This  leaves an open possibility that the nuclear region of the
Galaxy (within $\sim 10^\prime - 1^\circ$ around Sgr A$^\ast$) may
be somewhat different from the rest of the Galaxy, and, therefore,
processes of radiation there have an origin which differs from
other parts of the Galactic disk.

This region is known, indeed, to be peculiar in many respects:
\begin{itemize}
\item The plasma temperature in the GC
region is higher ($\sim 10$ keV) than in other parts of the
Galactic disk. Such a high plasma temperature is surprising, since
the gravitational potential in the GC region is no greater than
several hundred eV which is too small to bind the gas. The plasma
could not be gravitational confined and a very high amount of
energy is required to maintain the plasma outflow. This energy
supply cannot be produced by SN explosions and other more powerful
sources of energy are required to support the energy balance there
\citep{{sun1},koya1,muno} ;\\
\item GC region is a source of annihilation emission whose
origin is still enigmatic. It may be explained either by emission
of point sources like nova and supernova  (see e.g. \cite{derm})
or low mass X-ray binaries \citep{weid}, or by dark matter
annihilation \citep{sizun};\\
\item Intensive emission in X-ray iron lines is observed from the
Galactic center, which is often explained that the gas there was
exposed in the past by sources of intensive X-ray emission e.g.
from a supernova or from the Galactic nucleus \citep{sun1, koya1};\\
\item The temperature of molecular hydrogen ($H_2$) in GC is unusually high,
$T= 100- 200$~K. With the exception of Sgr B2 no embedded source
are observed inside clouds. Therefore, a global heating
mechanism is needed to explain the high gas temperature, e.g. by cosmic rays (see e.g. \citet{yuz});\\
\item The flux of VHE gamma-rays of unknown origin was discovered
by HESS in the direction of GC, which is supposed due to processes
nearby the central black hole \citep{aha}.
\end{itemize}
As one can   see, each of these observational phenomena can be
explained separately by completely different physical processes
which are not concerned
 with each others.

We assume, however, that these phenomena have common origin,
namely, they are consequences of star accretion onto the central
supermassive black hole. As our estimations showed the energy
produced by accretion in the form of relativistic and
subrelativistic particles is so high that it is inaccessible for
any other sources of energy in the Galaxy. Another important
characteristic of this process is that the energy erupted in GC is
conserved there for rather long time because of relatively slow
dissipation time of primary and secondary charged particles
generated by accretion processes. We developed this model in
series of papers  \citep{cheng1, cheng2,dog08,dog09}, in which we
interpreted the origin of annihilation emission and estimated the
flux of gamma-ray de-excitation lines from GC. This publication is
a continuation of these investigations.

Below we present a model of thermal and non-thermal hard X-ray
 emission from the GC which is supposed to be due to
 specific processes of accretion on the central black
 hole. We restrict our analysis by integral characteristics of
 this emission, its spatial variations  are  beyond the scope of this paper. We
 suppose to present such an analysis in our following publication. The goal of
 this paper is to demonstrate a principle possibility of
  of  X-ray production in the GC (thermal and non-thermal) by
  subrelativistic protons at the level observed by Suzaku.

\section{Proton Injection  and Plasma Heating.
The Origin of Thermal X-ray Emission from GC}
Processes of injection of subrelativistic protons by star
accretion were described in details in \citet{dog09}. Here we
remind only the main parameters of the process.

Every star capture by supermassive black holes releases a huge
energy which is several order of magnitude higher than produced by
a supernova explosion. The average frequency capture of one solar
mass stars by supermassive black holes is about $(1-10)\times
10^{-5}$year$^{-1}$ (see \cite{don} and \cite{syer}).  As it was
shown by
 \citet{ayal}  once passing the pericenter, the
star is tidally disrupted into a very long and dilute gas stream.
Approximately  $50-75$\% of the star matter was not accreted but
instead became unbounded. This unbounded mass receives an
additional angular momentum and escapes with velocities higher
than the orbital speed that corresponds to the energy per baryon
higher than
\begin{equation}
E_{esc} \sim \frac{2GM_{bh}m_{\rm p}}{R_T} \sim 5\times 10^7
M_6^{2/3}m_*^{1/3}r_*^{-1} \mbox{~eV}\,. \label{esc}
\end{equation}
For the mass of the black hole located in the center of our Galaxy
is about $(4.31 \pm 0.06)\times10^6$~M$_{\odot}$ (see,
\cite{guss})
 it follows that   the
average energy of escaped particles may be more than
$50-100$~MeV~nucleon$^{-1}$ when a  one-solar mass stars is captured
(see for details \cite{dog09}. Here $R_T$ is the capture radius of
a black hole given by
\begin{equation}
R_T \approx 1.4\times 10^{13}
M_6^{1/3}m_*^{-1/3}r_*\,\mbox{~cm}\,, \label{rt}
\end{equation}
where $m_*=M_*/M_\odot$, $M_6=M_{bh}/10^6M_\odot$,
$r_*=R_*/R_\odot$, $M_*$ and $R_*$ are the star mass and radius,
 $M_{bh}$ the mass of the black hole, and $M_\odot$ and $R_\odot$
 are the solar mass and radius.

For these parameters after every capture event about $10^{57}$
protons with energies $\sim 100$ MeV  escape into the surrounding
medium whose  temperature is $\sim 6.5$ keV and the uniform target
gas distribution was assumed \citep{koya1,muno,koya2}. The rate of
their energy losses is
\begin{equation}
\left(\frac{dE}{dt}\right)_i=-\frac{ 4\pi ne^4\ln\Lambda_1}{
m\mathrm{v}_{\rm{p}}}\,,
\end{equation}
where $\mathrm{v}_{\rm{p}}$ is the proton velocity, and $\ln\Lambda_1$ is
the Coulomb logarithm.  Since the lifetime of subrelativistic
protons
\begin{equation}
\tau_i=\int\limits_E\frac{dE}{(dE/dt)_i}\,,
\end{equation}
is about $10^{14}$ s.

Since the proton lifetime is much longer than the characteristic
time of star capture, the process of proton injection can be
considered as quasi-stationary with the rate of proton injection
in between from $Q=10^{45}$ to $10^{46}$protons s$^{-1}$ (or the
energy input $\dot{W}\sim 10^{42}$ erg s$^{-1}$).

The time-dependent spectrum of subrelativistic protons, $N({\bf
r},E,t)$ can be calculated from the equation
\begin{equation}\label{pr_state}
\frac{\partial N}{\partial t}  - \nabla D\nabla N +
\frac{\partial}{\partial E}\left( b(E) N\right) = Q(E,{\bf
r},t)\,,
\end{equation}
where  $D$ is the spatial diffusion coefficient of cosmic ray
protons whose average value in the was taken to be $D\simeq
10^{26}$ cm$^2$s$^{-1}$ in order to reproduce the Suzaku
data, $dE/dt \equiv b(E)$ is the rate of proton energy losses, and
$Q(E,t)$ is the rate of proton production by accretion, which can
be presented in the form
\begin{equation}
Q(E, {\bf r}, t) = \sum \limits_{k=0}Q_k(E)\delta(t -
t_k)\delta({\bf r})\,,
\end{equation}
where $t_k$ is the injection time.  The average time of star
capture in the Galaxy was taken to be $T\simeq 10^4$ years, then
$t_k=k\times T$.

The energy distribution of  erupted nuclei $Q_k(E)$ is taken as a
simple Gaussian distribution the energy distribution of these
erupted nuclei is taken as a simple Gaussian distribution ( in
order to avoid a simple delta function injection)
\begin{equation}\label{Qesc}
        Q_k(E)=\frac{N}{\sigma\sqrt{2\pi}} \exp\left[-\,\frac{(E-E_{esc})^2}{2\sigma^2}\right],
\end{equation}
where we take the width $\sigma=0.03E_{esc}$ with $E_{esc}\simeq
100$ MeV, and $N$ is total amount of particles ejected by one
stellar capture.

In the nonrelativistic case when the rate of energy losses can be
approximated as $(dE/dt)_i\simeq a/\sqrt{E}$, the solution
(\ref{pr_state}) can be presented as
\begin{equation}
f({\bf
r},E,t)=\sum\limits_{k=0}\frac{N_k\sqrt{E}}{\sigma\sqrt{2\pi}Y_k^{1/3}}
\frac{\exp\left[-\frac{\left(E_{esc}-Y_k^{2/3}\right)^2}{2\sigma^2}-\frac{{\bf
r}^2}{4D(t-t_k)}\right]}{ \left(4\pi D(t-t_k)\right)^{3/2}}\,,
\label{sol1}
\end{equation}
where
\begin{equation}
Y_k(t,E)=\left[\frac{3a}{2}(t-t_k)+E^{3/2}\right]\,.
\end{equation}

Below we use the following parameters of the model for
calculations:  $Q=2\times 10^{45}$protons s$^{-1}$, $E_{esc}=100$
MeV. The plasma temperature in the GC derived  from Suzaku
data and it equals about 6.5 keV , and the average plasma density
 there is about 0.2 cm$^{-3}$.

\begin{figure}[h]
\begin{center}
\FigureFile(80mm,80mm){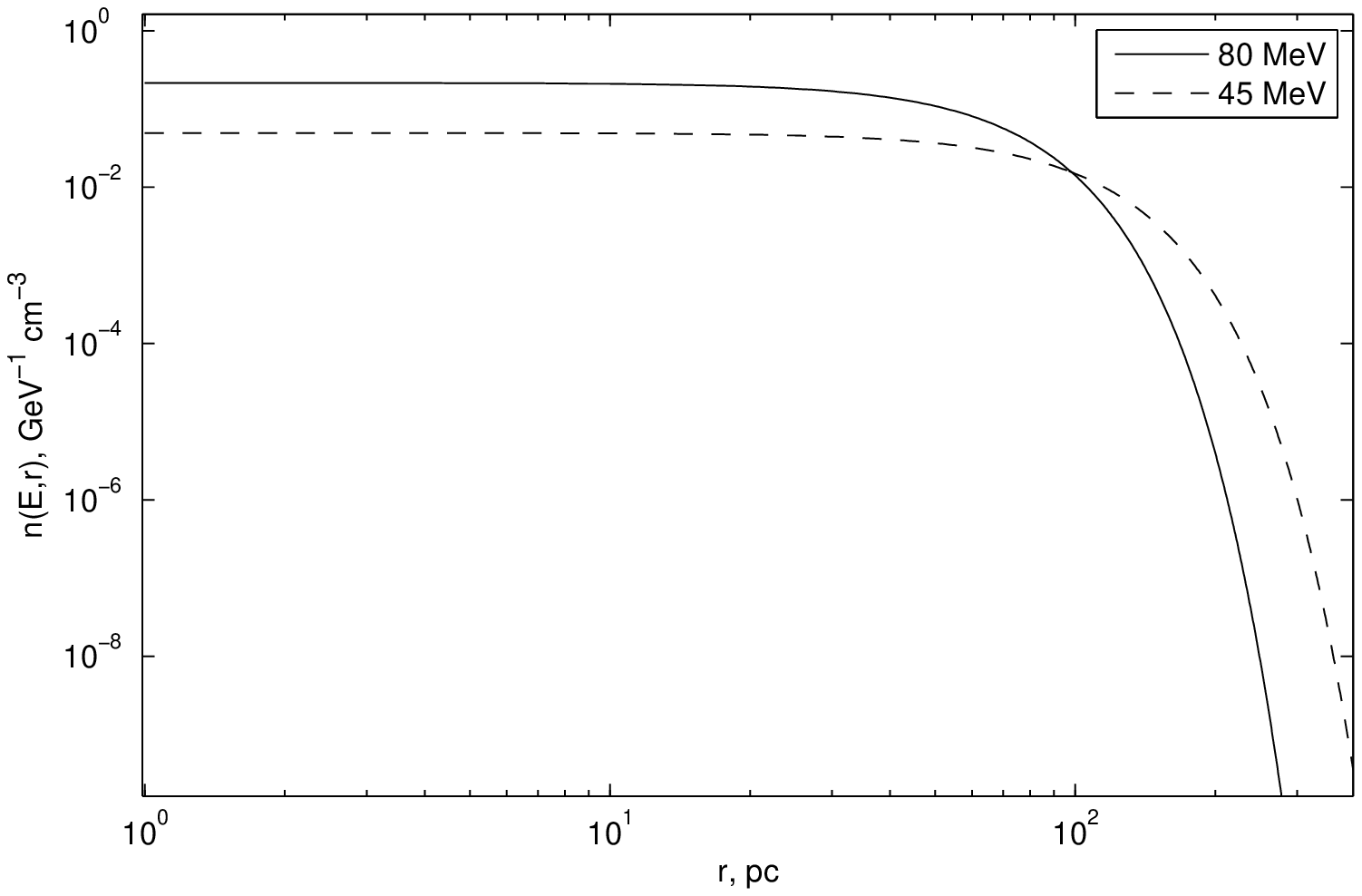}
\FigureFile(80mm,80mm){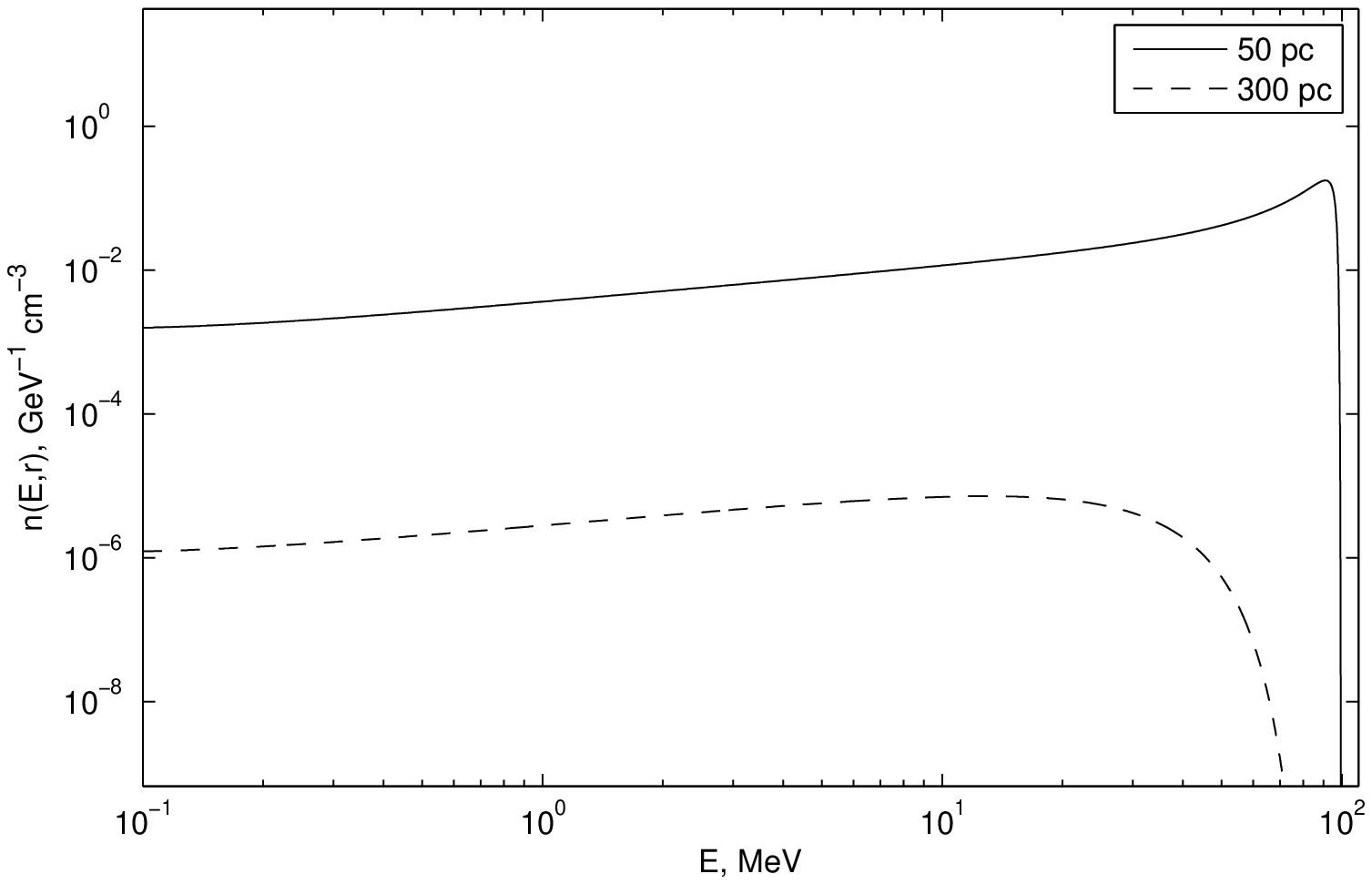}
\end{center}
\caption{ (a)Spatial distribution of protons. (b)Energy spectrum
of protons. }\label{two_dist}
\end{figure}

As am example we show in figure \ref{two_dist} spatial and energy
distributions of subrelativistic protons near the GC.

The energy releases in the Galactic center in the form of
subrelativistic protons may effectively heat the plasma there.
 Last Suzaku observations of \citet{koya2}
found a clear evidence for a hot plasma in the GC with the
diameter about 20 acrminutes (i.e $\sim 50-60$ pc). Total X-ray
flux from this region in the range 2 to 10 keV is $F_{\rm{X}}\sim 2\times
10^{36}$ erg s$^{-1}$, total energy of plasma in this region is
about $3\times 10^{52}$ erg. The temperature derived from the
continuous spectrum is about 15 keV but such a high value may be a
result of contribution from a non-thermal component of X-ray
emission. Indeed,  from the 6.9/6.7 keV line ratio \citet{koya2}
concluded that the spectrum in $5-11.5$~keV range was naturally
explained by a 6.5~keV-temperature plasma in collisional
ionization equilibrium.

The origin of the hot temperature plasma is unclear. Powerful
sources of heating are required. However, there are no evident
sources of energy in the Galactic central region. \citet{sun1},
\citet{koya1} and \citet{muno} concluded that the plasma in the
$1^\circ- 2^\circ$ radius central region can be heated up to the
observed temperature $T\sim 6- 10$~keV if the energy release there
is about $\dot{W}\sim10^{41}- 10^{42}$erg s$^{-1}$ that cannot be
provided by supernova explosions. Just this power can be supplied
by subrelativistic protons which loose their energy mainly by
ionization and thus  heating of background plasma.

Undoubtedly, this very crude estimate of temperature is severely
limited. In order to describe the temperature and gas distribution
in GC when the energy is sporadically released due to star capture
more sophisticated hydrodynamic and MHD calculations are required.
In this respect we mention recent results of \citet{breit} who
 simulated from 3D numerical calculations the dynamical structure of
interstellar medium in star formation region with respect to the
volume and mass fractions of the different ISM "phases". Due to
energy release from star explosions the medium  is strongly
nonuniform and turbulent. Compressed region of cold and dense
filaments co-exist with a hot and low density plasma. This reminds
the filamentary and non-uniform structure around GC.

\section{6.7 and 6.9 keV Iron Line Emission from GC}
As follows from conclusions of the previous section
subrelativistic protons heat effectively the background gas. We
search below whether they  generate also an excess of iron line
emission  in the  X-ray flux from the GC as it was expected from
in-situ accelerated non-relativistic electrons in the GRXE
spectrum  \citep{dog2,masai}.
  As follows
from \citet{dog4} subrelativistic protons effectively produce
shell vacancies that may be found in the line X-ray spectrum of
the Galactic center.

  Recent Suzaku
observations with high energy resolution clearly resolved several
 iron lines in the spectrum of hot plasma into individual peaks of FeI K$\alpha$ (6.4 keV),
 FeXXV K$\alpha$ (6.7~keV),
FeXXVI Ly$\alpha$ (6.9~keV),   FeXXV K$\beta$ (7.8~keV), FeXXVI
Ly$\beta$ + FeXXV K$\gamma$ (around $8.2-8.3$~keV), and FeXXVI
Ly$\gamma$ (8.7 keV)
 \citep{koya2,ebi2}. The  FeI K$\alpha$ emission is associated with
molecular clouds and therefore it requires a special analysis
which will be presented in other paper. The lines 6.7 keV and 6.9
keV,
 provide important information about the plasma
parameters since their intensities are proportional to the number
of  iron ions.

The He-like K-$\alpha$ emission consists of  emission lines of
different transitions, and it might even contains emissions from
Li-like ions in CCD energy resolutions.   The photon-weighted
centroid energy of the emission depends on the ionization process
(e.g. collisional or photo ionization) and the emission process
(e.g. thermal or non-thermal emission like charge exchange).  The
centroid energy derived from observation is consistent with that
the GC plasma is in the collisional ionization equilibrium as it
was claimed by \citet{koya2}.
 The ratio of
6.9/6.7 lines is proportional to abundances of FeXXV and FeXXVI
iron ions which  are functions of plasma temperature.  From the
best determined flux ratio of FeXXVI Ly$\alpha$ and FeXXV
K$\alpha$ lines  equaled $0.3-0.38$ in GC region \citet{koya2}
concluded that the electron temperature is
$\mathrm{kT}_{\mathrm{e}}=6.4\pm 0.2$ keV. \citet{yama} showed
that this ratio is almost constant in the Galactic disk but
increases in almost two times in the direction of GC that
indicates on higher temperatures in the Galactic center than in
the disk.

In principle, a flux of subrelativistic protons may provide
additional vacancies in iron ions that distorts the temperature
estimation obtained from the line ratio. In order to estimate the
effect from nonthermal protons one should accurately calculate 6.7
keV and 6.9 keV line intensities provided by thermal plasma and by
nonthermal particles.

A correct analysis of the ratio 6.7/6.9 lines was provided by
\citet{prokh} for the case of galaxy clusters who estimated a
mimic temperature excess due to nonthermal particles. Below we
present results of similar analysis for GC.

Taking into account both electron impact excitation and radiative
recombination, the line flux ratio of FeXXVI Ly$\alpha$/FeXXV
K$\alpha$  is given by
\begin{equation}
R=\frac {\xi_{\mathrm{FeXXVI}}Q^{1-2}_{\mathrm{FeXXVI}} +
\xi_{\mathrm{FeXXVII}}\alpha^{1-2}_{\mathrm{FeXXVI}}}
{\xi_{\mathrm{FeXXV}}Q^{1-2}_{\mathrm{FeXXV}} +
\xi_{\mathrm{FeXXVI}}\alpha^{1-2}_{\mathrm{FeXXV}}}\,,
\end{equation}
where the coefficients $Q^{1-2}_{\mathrm{FeXXV}}$ and
$Q^{1-2}_{\mathrm{FeXXVI}}$ describes processes of impact
excitation by thermal electron and subrelativistic protons for
FeXXV and FeXXVI respectively, $\alpha^{1-2}_{ \mathrm{FeXXV}}$
and $\alpha^{1-2}_{ \mathrm{FeXXVI}}$ are the rate coefficients
for the contribution from radiative recombination of the spectral
lines FeXXV (He-like triplet) and FeXXVI (H-like doublet)
respectively.

The rate coefficients are obtained by averaging the product of
cross section by particle velocity over the particle distribution
function. The ionic fractions of $\xi_{\mathrm{FeXXV}}$,
$\xi_{\mathrm{FeXXVI}}$ and $\xi_{\mathrm{FeXXVII}}$ are
calculated for the case of thermal plasma with a nonthermal
particle population. For corresponding references on cross
sections of ionization, recombination and impact excitation etc,
see \citet{prokh}.

\begin{figure}[h]
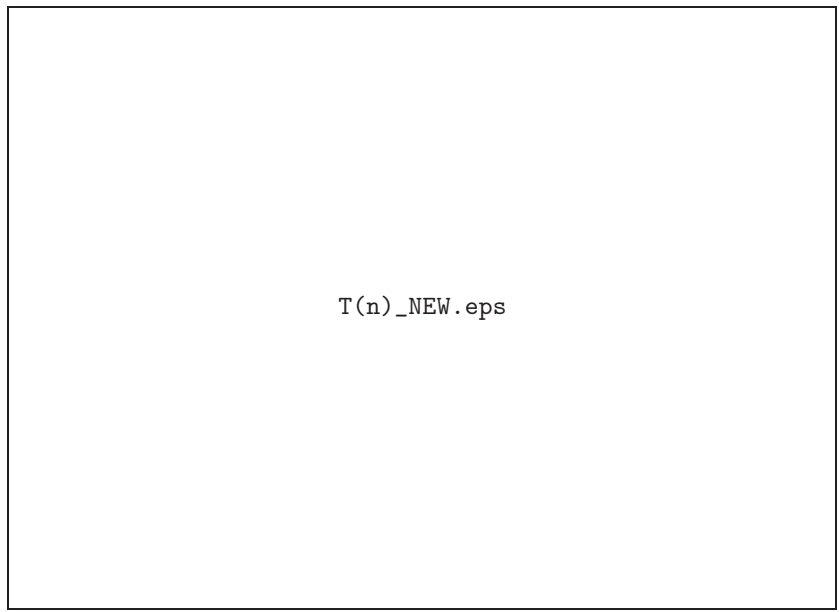

  \begin{center}
    \FigureFile(110mm,80mm){T(n)_NEW.eps}
  \end{center}
  \caption{Real temperature of plasma derived from the
  6.7/6.9 ratio as a function of the plasma density,
  when  ionization is provided also by subrelativistic protons.}\label{T(n)}
\end{figure}

In figure\ref{T(n)} we presented the temperature of plasma derived
from the observed 6.7/6.9 ratio which was fixed for the central
region and equaled 0.33. The contribution from subrelativistic
protons was calculated for the spectrum derived from equation
(\ref{pr_state}). We calculated the real plasma temperature for
the plasma density which changes in the range from 0.1 to 0.4
cm$^{-3}$.

We see that the contribution from subrelativistic protons is
negligible ($\leq 10$\%) and the calculated temperature is close
to the one
 derived by \citet{koya2} for the pure thermal case. The
reason is evident, the energy of lines are close to the plasma
temperature, therefore, these lines are mainly excited by thermal
particles.

We found that the value of the FeXXV K$\alpha$ line intensity for
the case with the nonthermal component differs from the pure
thermal case in a factor less than 1\% while intensity variations
of the FeXXVI Ly$\alpha$ line are more significant, they are at
the level of several percents . Just variations of FeXXVI
Ly$\alpha$ line leads to a higher "effective temperature" ($\sim
6.5$ keV) for the case of nonthermal particles contribution.

Our calculations show that "an X-ray signal" from subrelativistic
protons can be found at energies higher than 10 keV where the
influence of thermal emission is insignificant.

\section{Nonthermal Emission of Subrelativistic Protons from the GC}
Suzaku data of \citet{koya2} suggested  that the continuum flux
from the GC contained  an additional hard component. Up to
recently any direct confirmation of nonthermal emission at
energies above 10 keV has  been unavailable.  \citet{yuasa}
performed analysis of Suzaku data  and showed a prominent hard
X-ray emission in the range from 14 to 40~keV whose spectrum is a
power law with the spectral index ranging from 1.8 to 2.5. The
total luminosity of the power-law component from the central
region ($\mid l \mid<2^\circ,~\mid b \mid<0.5^\circ$) is $(4\pm
0.4)\times 10^{36}$ erg s$^{-1}$ . The spatial distribution of
hard X-rays correlates with  the distribution of hot plasma.

This spectrum can be represented by an exponentially cutoff power-law model,
\begin{equation}
f(E)=K(E/1~\rm{keV})^{-\Gamma}\exp(-E/E_c)\,,
\end{equation}
with $\Gamma$ and $E_c$ varying from region to region over 1.2 -
2.2 and $19 - 50$~keV, respectively.

\begin{figure}[h]
\begin{center}
\FigureFile(110mm,80mm){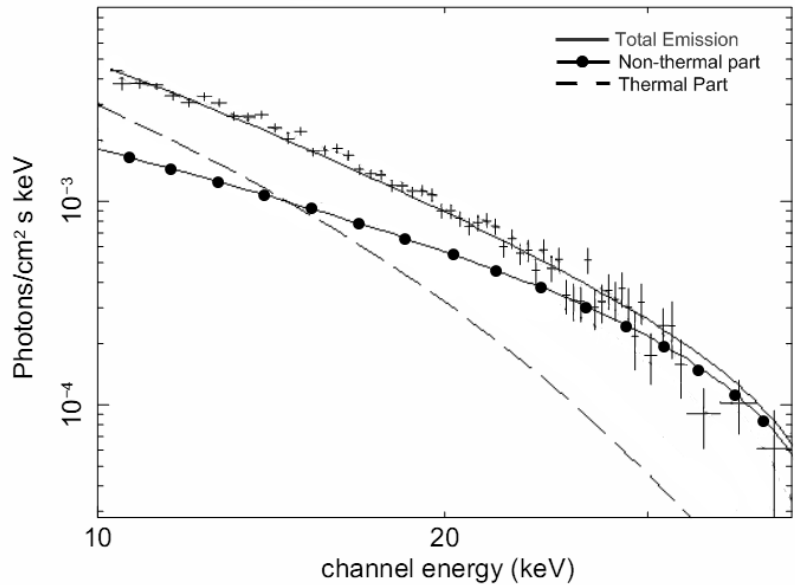}
\end{center}
\caption{The spectrum of inverse bremsstrahlung emission generated
by subrelativistic protons ($E_{\rm{esc}}=100 MeV$ was assumed,
see text) in the GC region (solid line with points) and the
de-convolved X-ray flux observed by the Suzaku HXD in the GC
region (crosses; Suzaku Observation ID 100027010). The thermal
X-ray emission for the temperature 6.4 keV is shown by the dashed
line and the total emission (thermal+inverse bremsstrahlung) is
shown by the solid line. Normalization factors are adjusted so
that the total model emission reproduces the observed HXD flux.}
\label{X-ray}
\end{figure}

Since the Hard X-ray Detector (HXD) onboard Suzaku is not an
imaging detector, \citet{yuasa} obtained the GC diffuse hard X-ray
spectra by subtracting contamination fluxes from known bright
X-ray point sources in its field of view ($34'\times34'$ FWHM).
They considered contributions from those point sources with fluxes
higher than $1.5\times10^{-11}~\ergcms$ (or $10^{-3}$ of a flux
from Crab Nebula) in the $14-40$~keV band, and subtracted them from
observed spectra with assumptions such as their spectral shapes
and fluxes are invariable during the observations. Resulting
energy spectrum of the GC hard X-ray emission observed by the HXD
after subtracting bright-point-source contaminations is shown in
figure \ref{X-ray} by crosses.

The residual spectrum still contains contributions from dimmer
point sources ($<1.5\times10^{-11}~\ergcms$ in the $14-40$~keV
band). Therefore before comparing the spectral shape and the
luminosity of our model with observed ones, we estimated the
remaining point source fluxes in the HXD spectra.

By integrating a luminosity function of X-ray point sources in the
GC region \footnote{"field" curve shown in figure 13 of
\cite{muno08} was used. In a small region around Sgr A*, rather
long exposure enabled high sensitivity and the luminosity function
was measured in the dimmer flux range. However we should use
"field" curve instead that of around Sgr A* because the HXD
observes wider region, i.e. $\sim$deg$\times$deg scale around the
Sgr A*.} over the luminosity range from $2\times10^{32}~\ergs$ to
$1\times10^{34}~\ergs$ (in the $0.5-8$~keV band; a distance of
8~kpc was assumed), we obtain a dim-point-source contaminating
flux of $\sim1.1\times10^{-15}~\ergcms$~arcmin$^{-2}$. In the
estimation, we took into account the effective solid angle of the
HXD/PIN of 1220 arcmin$^2$, and assumed that a spectral photon
index of the faint point sources is 1.5. The value is on the order
of 10\% of the HXD/PIN residual fluxes. If we further integrate
the luminosity function, extrapolating the measured one with the
same slope index down to $2\times10^{31}~\ergs$ and
$2\times10^{30}~\ergs$, the contaminating flux increase by 2.5 and
5.5 times ($\sim25\%$ and $\sim50\%$ of the HXD flux),
respectively. Since a precise qualitative treatment of the
contaminating point source flux is not a trivial procedure, we do
not deal them further, and compare our model spectra directly with
that of the HXD in the present analysis.

Interactions of subrelativistic protons with plasma result in
production of bremsstrahlung photons (inverse bremsstrahlung
radiation). Though the rate of these energy losses is negligible
in comparison with the above-mentioned Coulomb energy losses,
nevertheless, these losses generate emission in the energy range
higher than the thermal emission of background plasma and hence
can be observed. Subrelativistic protons  generate bremsstrahlung
photons  with characteristic energies about $E_{\rm{X}}< (m/M)E_{\rm{p}}$ where
$E_{\rm{p}}$ is the kinetic energy of protons and $m$ and $M$ are
 the masses of an electron and a proton. For the proton energies $E\leq
100$ MeV the bremsstrahlung radiation is in the range $E_{\rm{X}}< 55$
keV . The cross-section of inverse bremsstrahlung radiation is
\citep{haya}
\begin{equation}
  {{d\sigma_{\rm{br}}\over{dE_{\rm{X}}}}}={8\over 3}{Z^2}{{e^2}\over{\hbar c}}\left({{e^2}
  \over{m{c^2}}}\right)^2{{m{c^2}}\over{E^\prime}}{1\over{E_{\rm{X}}}}
\ln{{\left(\sqrt{E^\prime}+\sqrt{{E^\prime}-{E_{\rm{X}}}}\right)^2}\over{E_{\rm{X}}}}\,.
\label{sbr}
\end{equation}
Here  $E^\prime = (m/M)E_{\rm{p}}$. Then the total flux of inverse
bremsstrahlung emission from the GC can be calculated from
\begin{equation}
F_{\rm{X}}^{\rm{ib}}(E_{\rm{X}})=4\pi\int\limits_{E}dE\int\limits_{V_{\rm{GC}}}N_{\rm{p}}(E,{\bf
r},t){{d\sigma_{\rm{br}}\over{dE_{\rm{X}}}}}\mathrm{v}_{\rm{p}}n({\bf r})~d^3r\,,
\label{int_ib}
\end{equation}
where the  $V_{\rm{GC}}$ is the volume of emitting region.

The calculated   X-ray spectrum of inverse bremsstrahlung
radiation   is shown in figure \ref{X-ray} by the solid lines with
points. The GC hard X-ray spectrum observed with the Suzaku
 HXD shown by crosses \citep{yuasa} was deconvolved to
 a photon-spectrum to perform a direct comparison
 with our model spectrum of inverse bremsstrahlung.
In the deconvolution, as a rough approximation, we assumed a
$0.55^\circ\times 0.55^\circ$  spatially uniform emission (albeit
in reality the surface brightness of the emission has spatial
gradient centered on the GC). From spectral characteristics of the
Suzaku flux it follows that the energy $E_{esc}$ estimated by
equation (\ref{esc}) is no smaller than $85-100$~MeV, otherwise it
is impossible to reproduce the Suzaku data. The total spectrum
with the contribution of thermal emission is shown by the solid
line. The total flux of thermal component (without absorption) in
the energy range above 10 keV  is $F_{\rm{X}}^{\rm{th}}\simeq
2\times10^{36}$ erg s$^{-1}$. The inverse bremsstrahlung flux  for
the same energy range is $F_{\rm{X}}^{\rm{ib}}\simeq 3\times
10^{36}$ erg s$^{-1}$. We notice, however, that this estimate of
the nonthermal emission from the hot plasma in the GC is an upper
limit of the model, because continuum emission generated by these
protons in the molecular gas may contribute a significant part of
the total flux (see \cite{dog_p3}). Besides, if a part of line
emission comes from individual sources, the effect of the
non-thermal proton  becomes smaller.  Thus, the spectrum presented
in figure 3 is the worst case for the model.

The calculated flux of  inverse bremsstrahlung radiation from the
$0.55^\circ\times 0.55^\circ$ central region is weakly sensitive
to the average density of background plasma in the GC if the
spatial diffusion coefficient of protons is small enough. The
reason is that (as it follows from Eqs. (\ref{pr_state}) and
(\ref{int_ib})) the total flux of inverse bremsstrahlung radiation
can be estimated as
\begin{equation}
F_{\rm{ib}}\sim \bar{Q}\frac{\tau_i}{\tau_{\rm{ib}}}\,,
\end{equation}
where $\tau_i$ and $\tau_{\rm{ib}}$ are the characteristic times of
ionization and bremsstrahlung losses of protons, and $\bar{Q}$ is
the integrated power of proton sources. Since  both times are
proportional to the plasma density, the inverse bremsstrahlung
flux is almost independent of it.

We notice a significant discrepancy between the GC hard X-ray
emission and that of the GRXE. The last one can hardly be  due to
inverse bremsstrahlung of subrelativistic nuclei since in this
case the flux of carbon and oxygen de-excitation gamma-ray lines
in the range from 3 to 7 MeV is higher than the upper limit
measured by OSSE \citep{val}. However, if the hard X-ray
flux from the Galactic center is due to inverse bremsstrahlung of
protons with the derived spectrum, the flux of de-excitation lines
is still below the OSSE level \citep{dog09}. In this respect
common analysis of X-ray and gamma-ray data  is crucial for the
origin of hard X-rays from GC.

\section{Conclusion}
We analysed the origin of X-ray emission from GC assuming that it
is produced by subrelativistic protons generated by star accretion
on the central black hole. The average power of energy release
from accretion  is about $10^{42}$  erg s$^{-1}$ and the average
energy  of emitted protons is about 100~MeV. The energy of high
energy protons is transformed into plasma heating by ionization
losses. As derived by \citet{{sun1},koya1,muno} just this energy
release is necessary to heat the plasma  up to  temperatures about
$6-10$~keV, just as observed. Additional ionization of iron ions
by nonrelativistic protons can, in principle, violate the
ionization balance in GC providing an excess of FeXXVI ions that
increases the intensity of 6.9 keV iron line. However, as
numerical calculations show the excess due to ionization by
subrelativistic protons  is negligible for the 6.5 keV plasma
temperature. A more significant effect from subrelativistic
protons is expected in the X-ray range above 10 keV where
influence of thermal emission is insignificant. We show that the
inverse bremsstrahlung emission of protons in this energy range
may produce  a non-thermal X-ray flux. For the parameters of
accretion the inverse bremsstrahlung flux of protons is about
$3\times 10^{36}$ erg s$^{-1}$, i.e. about the flux observed by the
Suzaku from the GC in the $14-40$~keV band.

\vspace{5 mm} The authors are grateful to  K. Ebisawa, M. Ishida,
M. Revnivtsev, and S. Yamauchi for discussions and to the unknown
referee for useful comments.

 VAD and DOC were partly supported by the RFBR
grant 08-02-00170-a, the NSC-RFBR Joint Research Project No
95WFA0700088 and by the grant of a President of the Russian
Federation "Scientific School of Academician V.L.Ginzburg". KSC is
supported by a RGC grant of Hong Kong Government under HKU
7014/07P. CMK is supported in part by National Science Council,
Taiwan under the grant NSC-96-2112-M-008-014-MY3. A.~Bamba is
supported by JSPS Research Fellowship for Young Scientists
(19-1804).

\vspace{8 mm}


\end{document}